\documentclass[twoside,11pt]{article}
\usepackage{graphicx}
\usepackage{bm}
\usepackage{float}
\usepackage{tabularray}
\usepackage{booktabs}
\usepackage{lipsum}
\usepackage{cite}
\usepackage{appendix}
\usepackage[margin=1in]{geometry}
\usepackage[usenames]{color}
\usepackage{amsfonts, amssymb, amsmath, amsthm, multicol}
\usepackage[colorlinks=true, pdfstartview=FitV, linkcolor=blue,
citecolor=blue, urlcolor=blue]{hyperref}
\usepackage[usenames]{color}
\definecolor{Red}{rgb}{1,0.05,0}
\definecolor{Grn}{rgb}{0.1,0.7,0.1}
\definecolor{Blu}{rgb}{0.1,0.1,0.6}
\definecolor{Org}{rgb}{1,0.45,0}
\definecolor{Vio}{rgb}{0.6578,0,0.9478}
\definecolor{Mag}{rgb}{1,0.2,0.3}
\usepackage{authblk}
\usepackage{booktabs}
\usepackage{breqn}

\usepackage[section]{placeins}
\usepackage{graphicx}
\usepackage{amssymb}
\usepackage{epstopdf, epsfig}
\usepackage{color}
\usepackage{float}
\usepackage{amsmath}
\usepackage{tabularx}
\usepackage{bm}
\usepackage{accents}
\usepackage[super,sort&compress]{natbib}
\bibpunct{}{}{,}{s}{}{,}
\usepackage{array}
\usepackage{tabularx}
\usepackage{multirow,multicol, makecell}
\usepackage{pdflscape}
\usepackage{float}
\usepackage{breqn}
\usepackage{mathtools}
\usepackage{breqn}
\usepackage[flushleft]{threeparttable}
\usepackage{booktabs,caption}
\usepackage{footnote}
\usepackage{rotating}
\usepackage{lscape}
\makesavenoteenv{tabular}
\newcolumntype{C}[1]{>{\centering\arraybackslash}p{#1}}
\newcolumntype{L}[1]{>{\raggedright\arraybackslash}p{#1}}
\newcolumntype{M}[1]{>{\centering\arraybackslash}m{#1}}
\usepackage[mathlines,displaymath]{lineno}
\usepackage{lscape}
\usepackage{siunitx}
\usepackage{mathtools}
\usepackage[labelfont=bf]{caption}
\usepackage{hyperref}
\hypersetup{
    colorlinks,
    linkcolor={red!50!black},
    citecolor={blue!50!black},
    urlcolor={blue!80!black}
}

\usepackage{adjustbox}
\usepackage{xcolor}
\usepackage{soul}

\usepackage{enumitem}
\usepackage{lineno}

\numberwithin{rmk}{section}
\numberwithin{nt}{section}

\title{Droplets Suspended Beneath a Fiber Hub}

\author[1]{\textcolor{black}{Yi Zhang$^*$}}
\author[1]{\textcolor{black}{Zhao Pan}\thanks{To whom correspondence may be addressed: {yi.zhang,~zhao.pan}@uwaterloo.ca}}
\affil[1]{University of Waterloo, Department of Mechanical and Mechatronics Engineering, Waterloo, ON, Canada}

\vspace{-0.5 cm}
\date{\today}

\begin{document}
\maketitle 
\vspace{1cm}


\vspace{-2cm}
\begin{figure}[!h]
	\centering
	\includegraphics[width=0.6\linewidth]{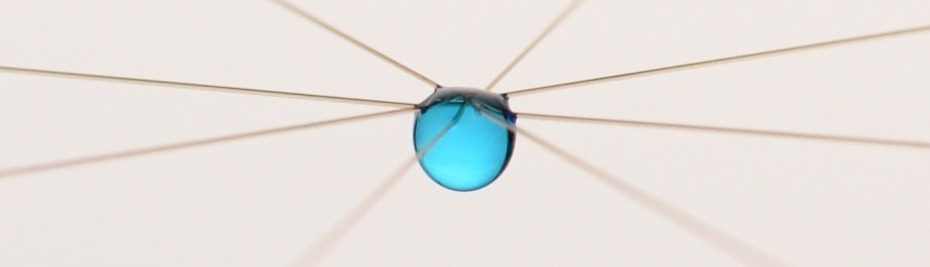}
\end{figure}

\vspace{-1cm}

\section*{Abstract}
Droplet-fiber interactions, prevalent in nature and widely applied across various engineering fields, have garnered significant research interest.
Many works have focused on the interactions between droplets and single or two fibers.
However, the wetting behavior of droplets, especially the maximum droplets that can be retained, on fiber hubs formed by many fibers is rarely studied. 
The current work explores the capability of fiber hubs to retain liquid droplets.
We develop analytical and semi-empirical models to predict the maximum droplet volume on a fiber hub, validating them against experimental data. 
The variation of maximum volume follows two distinct regimes as the fiber count increases, with a critical fiber number ($n^*$) marking the transition between them. 
In Regime I ($n\le n^*$), the volume increases with fiber number, and the stability of a droplet is dictated by the pinning of three-phase contact lines. 
In Regime II ($n>n^*$), the volume plateaus, with droplets under a fiber hub behaving similarly to those on a flat surface, where the stability is governed by Rayleigh-Taylor instability.

\noindent \textbf{Key words}: Droplet; Fiber hub; Maximum volume; Three-phase contact line; Stability

\section*{Introduction}
Interactions between droplets and fibers are commonly observed in nature, e.g., raindrops hanging from pine needles \cite{duprat2012wetting} and dew droplets on spider webs \cite{zheng2010directional}. They have inspired tremendous applications such as fog harvesting \cite{ju2012multi,hou2012water}, liquid-liquid separation \cite{rajgarhia2016separation}, and microfluidic devices \cite{gilet2009digital,weyer2015compound}. 
Droplet wetting on thin fiber---`quasi-one-dimensional' solid phase---is peculiar and significantly different from that on a flat surface (a two-dimensional interface) due to the geometry of thin cylindrical fibers \cite{mchale2002global}.
Therefore, the physics of droplets interacting with fibers is of great interest (see, for example, \cite{duprat2012wetting,pan2018drop,khattak2024directed} among others). 

On a single fiber, a droplet can adopt either a barrel or clamshell shape
\cite{mchale2002global}. 
Chou et al. \cite{chou2011equilibrium} developed a phase diagram for the droplet's state in various contact angles and liquid volumes, based on Surface Evolver simulations. 
For the droplet profile, Wu et al. \cite{wu2006droplet} modeled the morphology of microdroplets on a single fiber, assuming negligible gravity effect. 
In contrast, Mei et~al.~\cite{mei2013gravitational} examined the influence of gravity on droplet profiles and develop a model for the three-phase contact line (TCL) of barrel-shaped droplets. 
In addition to droplet profile, the maximum volume of droplets that can be retained by a horizontal fiber \cite{lorenceau2004capturing}, a bent fiber \cite{pan2018drop}, a patterned fiber with periodic spindle knots \cite{hou2012water} have also been explored, and the force to detach a droplet from a horizontal fiber is quantified \cite{farhan2018universal}. 
Furthermore, there are also studies focusing on the interactions between dynamic droplets with a horizontal rigid \cite{lorenceau2004capturing} or flexible fiber \cite{dressaire2016drop}, and the behaviors of droplets sliding along a tilted or vertical fiber \cite{gilet2010droplets,leonard2023droplets}.

On two fibers, either in parallel or crossed configuration, droplets behave differently from those on a single fiber. 
Protiere et al.~\cite{protiere2013wetting} reported that droplets on two parallel fibers can adopt one of the two states: a liquid drop or a column. 
A phase diagram for these two states and the transition criteria are developed. 
Wang and Schiller \cite{wang2021hysteresis} further specified the drop state on two parallel fibers into a barrel-shaped droplet and a droplet bridge. 
A bistable region corresponding to the transition between these two states is identified based on lattice Boltzmann simulations.
Sauret et al.~\cite{sauret2014wetting} presented that on two crossed fibers, liquid can fully spread into a column or form a drop either in the crossed node or at the end of the column. 
Moreover, for the two crossed fibers, Gilet et al. \cite{gilet2009digital} and Weyer et al. \cite{weyer2015compound} studied the dynamics of droplets sliding over the intersection, where residual droplets form due to the pinning effect.
This pinning effect and controlled residual droplets can be exploited to develop microfluidic devices for drug encapsulation, medical diagnosis, etc. 
More recently, Khattak et al. \cite{khattak2024directed} reported the directed movement of droplets between two converging fibers towards the apex, demonstrating a novel system for long-distance droplet transport.

For multiple fibers, Princen \cite{princen1970capillary} modeled the shape of the liquid column formed between horizontally aligned fiber arrays. 
The dynamics of droplets penetrating and sliding along vertically aligned fiber arrays were studied by Cardin et al. {\cite{cardin2023droplet}} and Leonard et al. {\cite{leonard2023droplets}}, respectively.
However, the wetting behavior of droplets on many intersecting fibers---particularly the maximum volume a droplet can retain at the fibers' intersections---has been rarely reported in the literature.
Droplets suspended from the hub of intersecting fibers, such as the junctions of spider webs \cite{qin2015structural} and the roots of plants in some hydroponic setups~\cite{lakhiar2018modern}, are also commonly observed in nature and industry.
Gaining a deeper understanding of these phenomena is not only of academic interest but enables industrial applications.
This paper aims to investigate how fiber hubs---structures composed of multiple horizontal fibers intersecting at a single point---with varying numbers of fibers retain liquid droplets against gravity.

The rest of this paper is structured as follows. 
\nameref{sec: methods} introduces the experimental methods for characterizing the maximum droplets hanging from fiber hubs with various number of fibers. 
Three different liquids, spanning a wide range of surface tensions, yet all exhibiting good wettability on the fiber surface, are employed.
\nameref{sec: Results} Section first presents the experimental results and then develops models predicting the maximum droplet volume. 
Last, the model is validated against experimental data before concluding the work in \nameref{sec: Conclusion} Section.

\section*{Experimental Section}
\label{sec: methods} 
\begin{figure}[!h]
    \centering
    \includegraphics[width=1\linewidth]{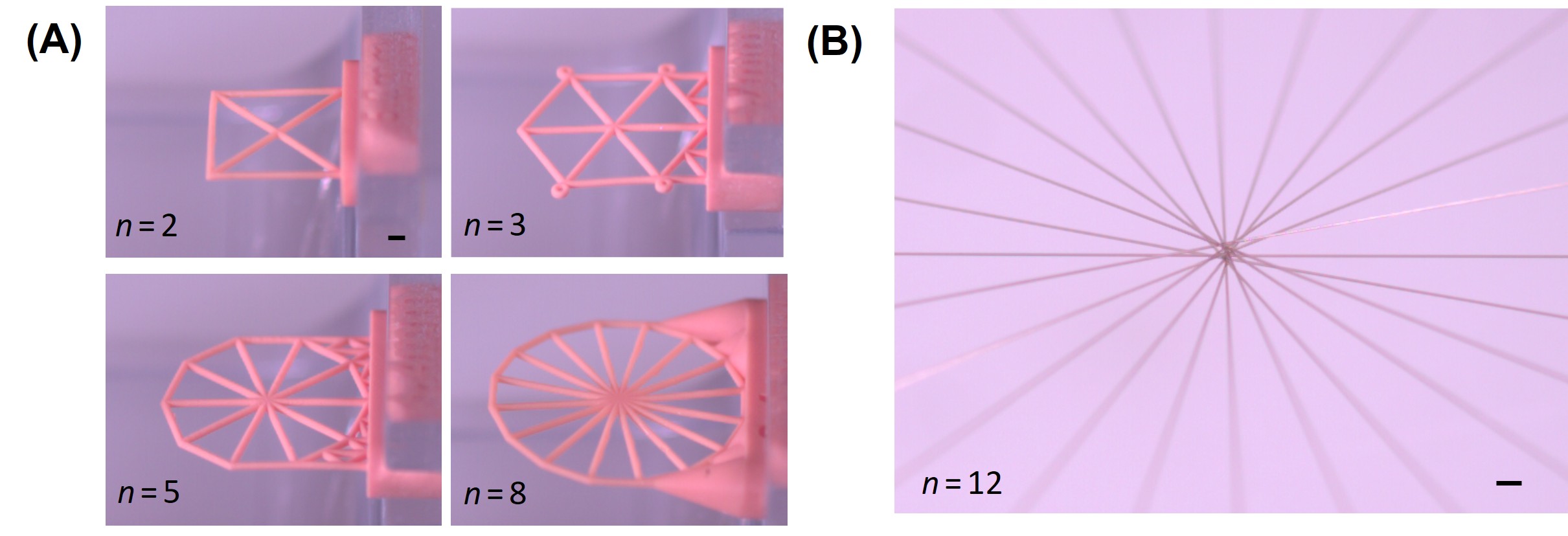}
    \caption{Photos of (A) the 3D-printed fiber hubs composed of 0.5~mm-diameter resin fibers and (B) the stacked fiber hubs consisting of 0.1~mm-diameter nylon fibers. Scale bars indicate 1~mm.}
    \label{fig:1}
\end{figure}

Figure~\ref{fig:1} presents two types of fiber hubs we construct to conduct experiments.
The first type, illustrated in Figure~\ref{fig:1}(A), composed of 0.5~mm diameter resin fibers in varying numbers ($n$, ranging from 1 to 96), is manufactured by 3D printing (EnvisionTEC Pro XL, Germany). 
The second type, shown in Figure~\ref{fig:1}(B), is constructed with a circular rigid frame, through which varying numbers~($n$) of 0.1 mm-diameter nylon fibers are stretched, meeting at the center to form a fiber hub.

Three solutions, i.e., \SI{0.01}{M} and \SI{0.001}{M} sodium dodecyl sulfate (SDS) solutions, and a mixture of 25~wt\% glycerol in water, are used. 
The surface tensions ($\gamma$) of the three solutions are measured to be \SI{34}{mN/m} (\SI{0.01}{M} SDS), \SI{44}{mN/m} (\SI{0.001}{M} SDS), and \SI{67}{mN/m} (glycerol-water mixture) using the pendant drop method via a contact angle goniometer (Ossila Ltd., UK). 
The densities~($\rho$) of the three solutions are \SI{1.00e3}{kg/\cubic\meter} for the two SDS solutions and \SI{1.06e3}{kg/\cubic\meter} for the glycerol mixture. On a flat resin (the same material with that of the 3D-printed fiber) surface, the contact angles ($\theta$) are ${\sim}6$$^{\circ}$ and  ${\sim}48^{\circ}$ for the \SI{0.01}{M} and \SI{0.001}{M} SDS solutions, respectively. The contact angle of the glycerol-water mixture on a nylon flat surface is ${\sim}30^{\circ}$.
In this work, we restrict our study to small contact angles ($<50^{\circ}$), as high contact angles may prevent the liquid from uniformly wetting all the fibers and hanging beneath the fiber hub. In such cases, droplets tend to become asymmetric and often stay on the top of the fiber hub.

Before experiments, the fiber hubs are rinsed with flowing deionized water, air-dried, and then held horizontally by a support. 
The solution is first loaded into a syringe, which is connected to a stainless steel needle (gauge 26) positioning just above the fiber hub.
The syringe is then driven by a syringe pump (Chemyx Fusion 200, USA) at a low flow rate of \SI{0.5}{\micro\liter/s}, delivering the solution to the fiber hub.
A droplet is formed underneath the fiber hub and its volume gradually increases until it falls, at which point the maximum volume is determined by multiplying the flow rate by the pumping time. 
The shape of the droplet (the front view) before dropping off is recorded using a high-speed camera (Kron Technologies, Canada). The top view that reveals the three-phase contact lines (TCL) is recorded by a stereo microscope with a magnification ratio of 40$\times$ (AmScope, USA).

\section*{Results and Discussion}
\label{sec: Results}
\subsection*{Experimental Results} \label{subsec: Experimental results}
\begin{figure}[!h]
    \centering
    \includegraphics[width=1\linewidth]{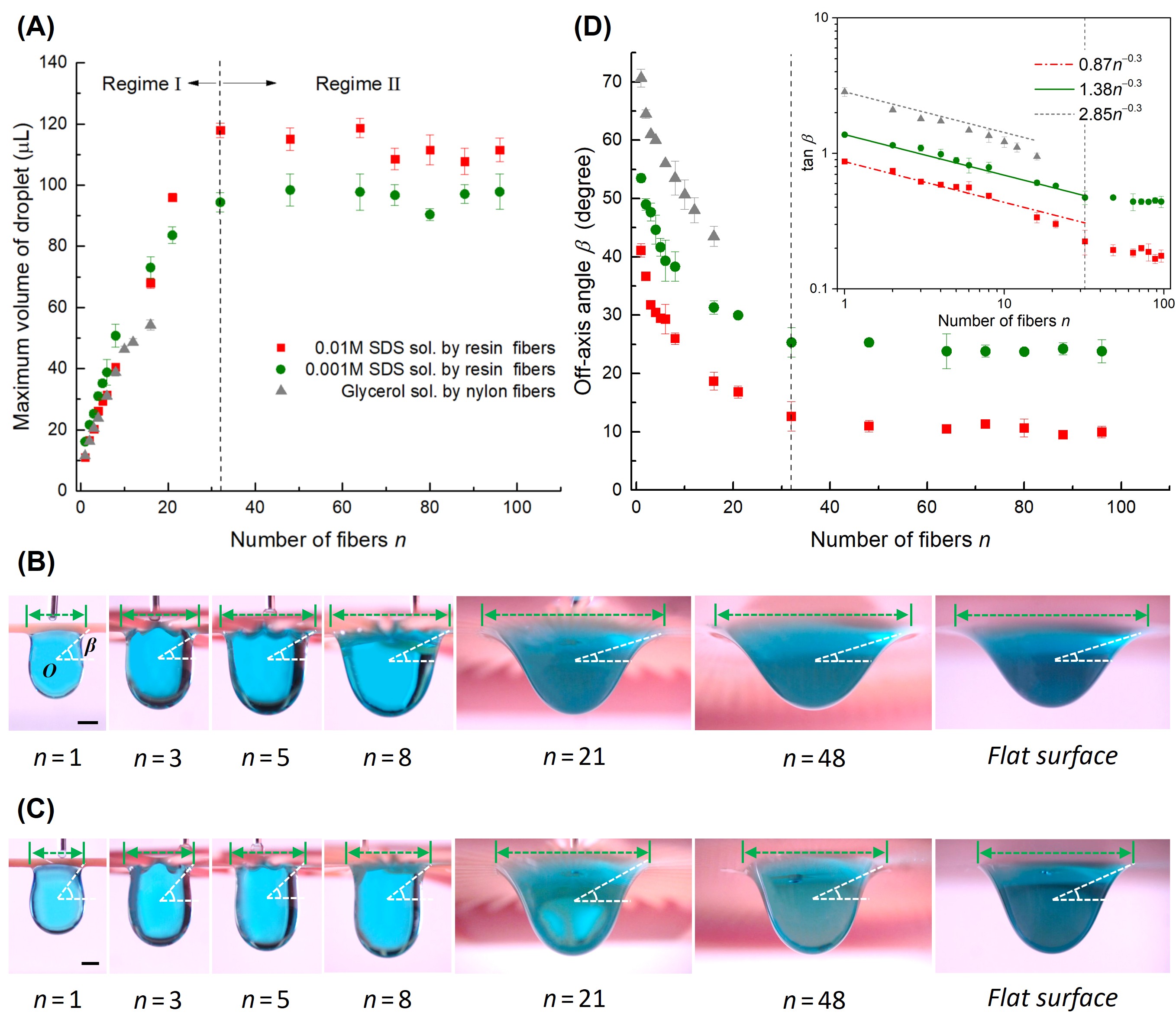}
    \caption{(A) Experimental data for the maximum droplet volume with increasing number of fibers (the vertical dashed line represents the critical fiber count $n^*= 32$, as also shown in (D); error bar indicates the standard deviation from the mean of 3--5 measurements; for $n\geq48$, the maximum droplet volume of \SI{0.01}{M} SDS solution adopts the volume of the bulk droplet, see Electronic Support Information (ESI) Section 1). Photos of maximum droplets for $n=$ 1, 3, 5, 8, 21, 48, and flat surface for (B) 0.01 M and (C) 0.001 M SDS solutions (scale bar indicates \SI{1}{mm}; the arrowed line segments indicate the nominal diameter of the wetted area). (D) Variation of off-axis angle $\beta$ with number of fibers $n$ (shares legend with (A)).}
    \label{fig:2}
\end{figure}

Figure~\ref{fig:2}(A) illustrates the maximum volume of droplets that can be retained beneath fiber hubs as the number of fibers increases.
As indicated by the gray triangle symbols, the maximum volume of the glycerol-water mixture droplets increases with the number of nylon fibers up to $n=16$. 
For $n>16$, despite the fibers being as thin as \SI{0.1}{mm} in diameter, the stacking height of fibers at the junction reaches ${\sim}$\SI{2}{mm}, which is not negligible for droplets of tens of microliters (radius of ${\sim}$\SI{2}{mm}). 
In this case, fibers cannot contribute equally to support droplets, and the nylon fiber hubs with $n>16$ will not be discussed in this paper. 
The red square and green circle symbols denote the results for the \SI{0.01}{M} and \SI{0.001}{M} SDS solutions, respectively, hanging underneath the 3D-printed fiber hubs. 
The maximum volume of the droplet first increases with the number of fibers for $n\leq32$ and then plateaus for $n>32$ for both solutions. 
Based on these observations, we divide the data in Figure~\ref{fig:2}(A) into regimes~I and~II by a critical fiber count $n^*= 32$, which will be discussed shortly. 

Figure~\ref{fig:2}(B) and~(C) display images of the droplets at their maximum volume suspended from fiber hubs with $n= 1, 3, 5, 8, 21, 48$, and a flat surface for the \SI{0.01}{M} and \SI{0.001}{M} SDS solutions, respectively. 
For $n\leq n^*$, the shape of the droplet varies with the number of fibers.  
The spreading of droplets along the fiber is enhanced while $n$ increases (indicated by the escalated length of the green dashed line segments for $n =1, 3, 5, 8, 21$),  while the droplet's center of mass (COM) remains presumably unchanged. 
This results in a reduction of the off-axis angle $\beta$, which is defined as the angle between the horizontal and the line connecting the COM (indicated by $O$ in Figure~\ref{fig:2}(B)) of the droplet and the three-phase point where the fiber exits the droplet.
The method for identifying the location of COM can be found in ESI Section 2.
For $n>n^*$ (e.g., $n=48$ in (B) and (C)), the droplet shape closely resembles that of a droplet suspended beneath a flat surface. 
Figure~\ref{fig:2}(D) shows the variation of $\beta$ with fiber numbers. 
For both \SI{0.01}{M} and \SI{0.001}{M} SDS solutions, $\beta$ first decreases and then plateaus as $n$ increases.

\begin{figure}[!h]
    \centering\includegraphics[width=0.8\linewidth]{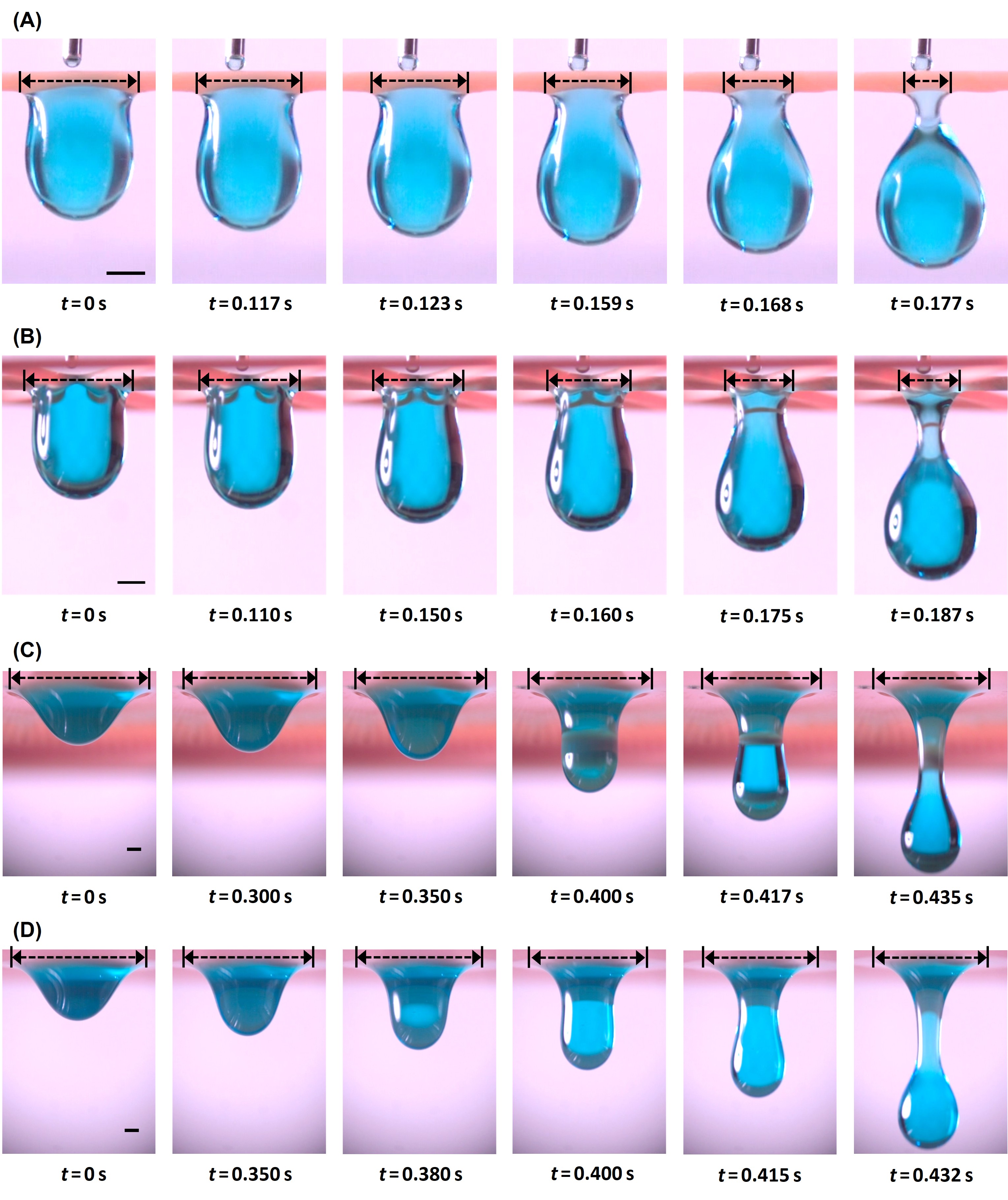}
    \caption{Droplets lose stability and fall down in regime I (A) \(n=1\) and (B) \(n=3\), and regime II (C) \(n=48\) and (D) flat surface (scale bar indicates \SI{1}{mm}; the arrowed line segments indicate the nominal diameter of the wetted area, as well as the movement of three-phase contact points).}
    \label{fig:3}
\end{figure}

Figure~\ref{fig:3}(A-B) and (C) show typical processes of droplets detaching from fiber hubs in regimes I and II, respectively, while Figure~\ref{fig:3}(D) presents the detachment of a droplet from a flat surface.
In regime I ($n\leq n^*$), the stability of droplets is dictated by the pinning of the TCLs. 
When the droplet reaches its maximum volume, external perturbations can trigger de-pinning of TCLs and the droplet drops off. 
The de-pinned TCLs are evident by observing the shrinking line segments as shown in Figure~\ref{fig:3}(A-B). 
The three-phase contact points move inwards along the fibers and then the droplet detaches due to a pinch-off close to the fiber hub. 
In regime II ($n>n^*$), the mechanism of droplet dropping off from a fiber hub is similar to that on flat surfaces.
When a droplet exceeds its critical volume, the TCL overall remains pinned, while the detachment is instead mainly driven by necking, which is similar to the observation for Rayleigh-Taylor instability \cite{taylor1950instability}.
As shown in Figure~\ref{fig:3}(C-D), the arrowed line segments indicating the pinned TCLs remain almost unchanged as the droplet begins to elongate downward.
A neck then forms, progressively separating the droplet into two parts and ultimately leading to drop-off. Note, the pinch off location in this case is much lower than that observed in the experiments for regime I. 

\subsection*{Modeling}
\label{sec: modeling}

Based on the experimental data and observations presented in \nameref{subsec: Experimental results} Section, models for the maximum volume of droplets for the two regimes are established as follows.

\subsubsection*{Regime I}

\begin{figure}[!h]
    \centering\includegraphics[width=1\linewidth]{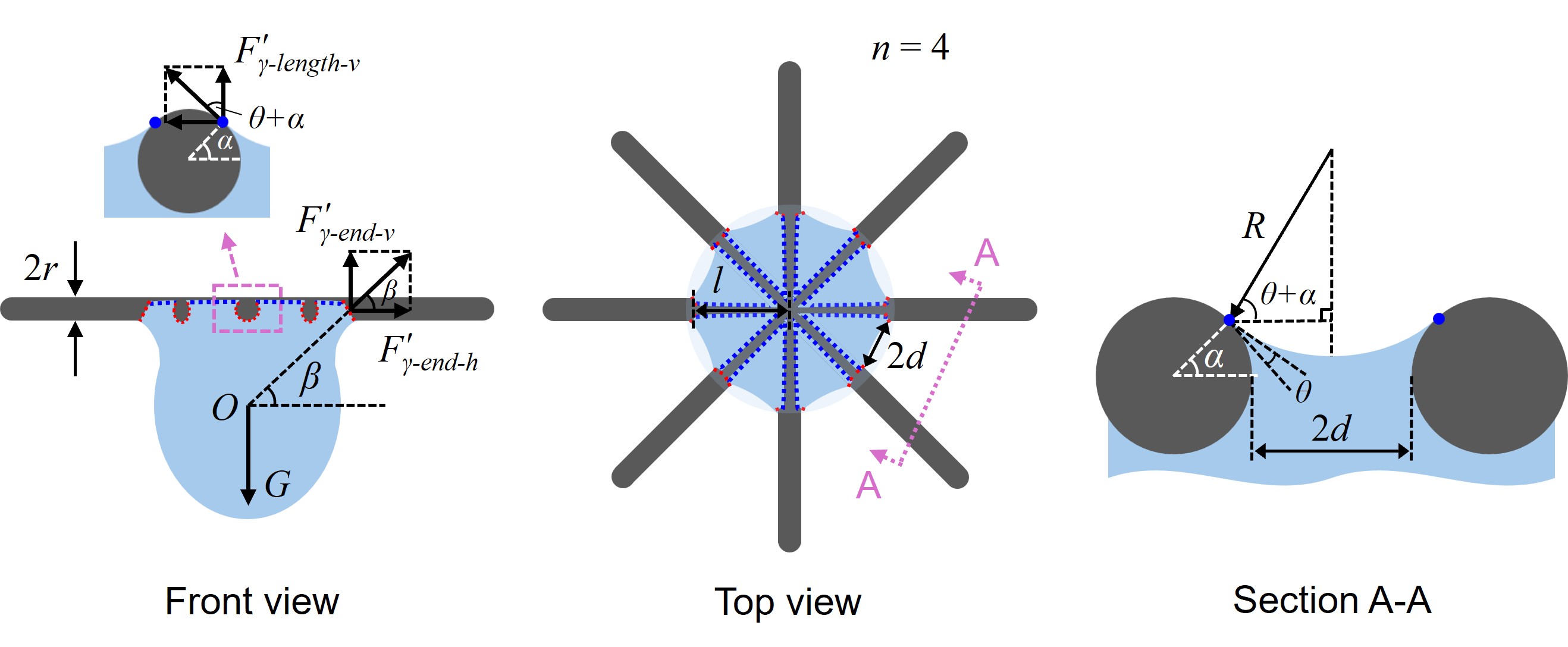}
    \caption{Schematic of a droplet hanging under a fiber hub. $F'_{\gamma\textrm{-}end\textrm{-}v}$ and $F'_{\gamma\textrm{-}end\textrm{-}h}$ are the vertical and horizontal components, respectively, of the capillary force acting at the ends of the wetted portion of each fiber (red dotted lines), $F'_{\gamma\textrm{-}length\textrm{-}v}$ is the vertical component of the capillary force acting on the horizontal portion of each fiber (blue dotted lines), $G$ the weight of droplet, $2d$ the distance between fibers at the end of the liquid bridge, $l$ the exposed length of TCL on a half fiber,  $n$ the number of fiber, $r$ the radius of fiber, $R$ the radius of meniscus of the liquid bridge between fibers, $\alpha$ the position angle of TCL along a fiber, $\beta$ the off-axis angle, $\theta$ the contact angle.}
    \label{fig:4}
\end{figure}

In regime I, where the maximum volume of droplet increases with the number of fibers, the force balance of the droplet is presented in Figure \ref{fig:4} and can be expressed as 
\begin{equation}
\label{eq:1}
 G = \Omega \rho g = F_{\gamma\textrm{-}end\textrm{-}v} + F_{\gamma\textrm{-}length\textrm{-}v},
\end{equation}
where $G$ and $\Omega$ represent the weight and volume of droplet, respectively; $\rho$ and $g$ are the density of liquid and gravitational constant, respectively; $F_{\gamma\textrm{-}end\textrm{-}v}$ and $F_{\gamma\textrm{-}length\textrm{-}v}$ denote the vertical components of the capillary forces acting at the ends of the wetted portion of the fibers and on the horizontal portion of the fibers, respectively. 
It should be noted that the Laplace pressure of the liquid-air interface is not considered in \eqref{eq:1} and $F_{\gamma\textrm{-}length\textrm{-}v}$ vanishes when the droplet fully wraps the fibers.

At the maximum volume, the TCLs are assumed to consist of partial circles at the ends of the wetted portion of the fibers, connecting by horizontal lines along the fiber. 
Therefore, the total length of the TCLs is 
$4r\left(\pi/2+\alpha\right)$
and $4l$, respectively, for each fiber. 
$r$ denotes the fiber radius, $\alpha$ the position angle of TCL along fiber, and  $l$ the exposed length of TCL on a half fiber (see Figure~\ref{fig:4} for illustration). 
The vertical component of the capillary force for a single fiber at the ends of the wetted portion of the fiber, $F'_{\gamma\textrm{-}end\textrm{-}v}$, can be calculated using \eqref{eq:2} based on the model developed by Pan et al.~\cite{pan2018drop} 
\begin{equation}\label{eq:2}F'_{\gamma\textrm{-}end\textrm{-}v}=F'_{\gamma\textrm{-}end\textrm{-}h}\tan\beta\approx4 r\left(\frac{\pi}{2}+\alpha\right)\gamma\cos\theta\tan\beta,
\end{equation}
where $\theta$ is the contact angle and $\beta$ is the off-axis angle. 
In \eqref{eq:2}, the droplet is assumed to be spherical~\cite{pan2018drop}.
With this assumption, the significance of $\beta$ can be understood in two ways.
For free energy analysis, $\beta$ specifies the movement of three-phase contact points when the COM of the droplet is perturbed. 
For force analysis, $\beta$ indicates the direction of the capillary force acting at the ends of the wetted portion of the fibers.
Assuming that the droplet uniformly wets each fiber of a fiber hub, $F_{\gamma\textrm{-}end\textrm{-}v}$ for the $n$ fibers can be computed as 
\begin{equation}\label{eq:3}
F_{\gamma\textrm{-}end\textrm{-}v}=nF'_{\gamma\textrm{-}end\textrm{-}v}.
\end{equation}
Similarly, $F_{\gamma\textrm{-}length\textrm{-}v}$ for the $n$ fibers in \eqref{eq:1} can be calculated based on the geometry shown in the front view of Figure~\ref{fig:4} as
\begin{equation}\label{eq:4}
F_{\gamma\textrm{-}length\textrm{-}v}=nF'_{\gamma\textrm{-}length\textrm{-}v}\approx4nl\gamma\cos{(\theta+\alpha)}.
\end{equation}
Substituting \eqref{eq:3} and \eqref{eq:4} into \eqref{eq:1} gives the maximum volume of the droplet ($\Omega_{\text{I}}$):
\begin{equation}\label{eq:5}
\Omega_{\text{I}} \approx 4n \lambda^2 \left[ r\cos\theta\tan\beta\left(\frac{\pi}{2}+\alpha\right)+l\cos(\theta+\alpha)\right],
\end{equation}
where $\lambda =\sqrt{\frac{\gamma}{\rho g}}$ is the capillary length of the liquid. 
 
Taking $\Omega_\lambda = \frac{4}{3}\pi\lambda^3$ as characteristic volume of a spherical droplet whose radius is the capillary length, normalizing \eqref{eq:5} leads to the non-dimensional maximum volume for regime~I: 
\begin{equation}\label{eq:6}
\Omega^*_{\text{I}} (n) = \frac{\Omega_{\text{I}}}{\Omega_\lambda}
 \approx n \frac{3}{\pi} \left[r^*\cos\theta\tan\beta\left(\frac{\pi}{2}+\alpha\right)+l^*\cos{(\theta+\alpha)}\right], \quad \text{for}~n \le n^*,
\end{equation}
where $r^* = r/\lambda$ and $l^* = l/\lambda$ are non-dimensional fiber radius and the exposed length, respectively.

By investigating the relationship between $\beta$ and $n$, as shown in the inset of Figure~\ref{fig:2}(D), we observe a linear correlation between $\log(\tan\beta)$ and $\log(n)$ in regime I. 
Notably, the slopes of the data for various solutions and fibers are nearly identical, approximately $-0.3$. 
This suggests a potential semi-empirical model for $\beta$ as a function of $n$:
\begin{equation}\label{eq:7}
\log(\tan\beta)\approx -0.3\log(n)+\log(C),
\end{equation}
where $C=\tan\beta_{n=1}$ and $\beta_{n=1}$ is the value of $\beta$ for $n=1$ for a given pair of liquid and fiber.
In our experiments, $C$ is calculated to be 0.87, 1.38, and 2.85 for \SI{0.01}{M} SDS, \SI{0.001}{M} SDS, and 25~wt\% glycerol solutions, respectively. 
Rearranging \eqref{eq:7} gives
\begin{equation}\label{eq:8}
\tan\beta\approx C n^{-0.3},
\end{equation}

Furthermore, the geometry in section A-A and the top view of Figure \ref{fig:4} gives 
\begin{equation}\label{eq:9}
\cos{(\theta+\alpha)}=\frac{r+d-r\cos{\alpha}}{R}=\frac{l}{R}\sin{(\frac{\pi}{2n})},
\end{equation}
which is also available in exiting literature (see e.g., \cite{princen1970capillary,sauret2014wetting}).
The rightmost term in \eqref{eq:9} is found to be $\lesssim 0.1$, see ESI Section 3 for details, leading to $\cos{(\theta+\alpha)}\lesssim 0.1$.
Accordingly, an approximation of $\theta+\alpha\approx\pi/2$ is made to further simplify~\eqref{eq:6}. 
Substituting \eqref{eq:8} and $\theta+\alpha\approx\pi/2$ into \eqref{eq:6} gives
\begin{equation}\label{eq:10}
\Omega^*_{\text{I}}(n) \approx 3C n^{0.7}  r^*\left(1-\frac{\theta}{\pi}\right)\cos\theta, \quad \text{for}~n \le n^*.
\end{equation}

Assuming $\beta$ remains constant at $\beta_{n=1}$, meaning that $\Omega^*_{\text{I}}(n) \approx n \Omega^*_{\text{I}}(n=1)$,
an asymptote of \eqref{eq:10} in the limit of $n\to1$ is given as
\begin{equation}\label{eq:11}
\Omega^*_{\text{I}}(n) \approx n \Omega^*_{\text{I}}(n=1) = 3Cn r^*\left(1-\frac{\theta}{\pi}\right)\cos\theta.
\end{equation}
Physically, \eqref{eq:11} evaluates $\Omega^*_{\text{I}}(n)$ for multiple fibers by scaling $\Omega^*_{\text{I}}(n=1)$ for a single fiber linearly with $n$.
This scaling ($\Omega^*_{\text{I}}(n)\sim n$) may align with naive intuition. 
However, as shown in our experiments and will be soon discussed in detail, it is inaccurate and overestimates the growth rate of the maximum volume with respect to $n$, as it ignored the fact that $\beta$ decreases with $n$ in regime I. 
As $n\to \infty$, $\Omega^*_{\text{I}}$ in \eqref{eq:6}, \eqref{eq:10}, or \eqref{eq:11} blows up, which is nonphysical. 
A valid model for $n\to \infty$ is needed, which will be presented immediately in the following section.

\subsubsection*{Regime II}
\label{sec:Regime II mdoel}
The maximum volume of droplet held at the center of the fiber hub plateaus with increasing $n$, meaning that the $\Omega^*_{\text{II}}$ is roughly a constant with respect to $n$ (Figure~\ref{fig:2}(A)).
This observation is expected as it is natural to view a dense fiber hub as a flat surface in the limit of $n\to\infty$, and the maximum volume of the liquid suspended on the ceiling due to gravity is a classic problem of Rayleigh–Taylor instability \cite{taylor1950instability,padday1973stability}. 
The non-dimensional maximum volume of the droplet hung under a flat surface ($\bar{\Omega}^*_\theta$, normalized by $\Omega_\lambda$) before dropping off is a function of the contact angle ($\theta$) of the surface alone. 
Thus, we expect the maximum volume of the droplet held at the center of a fiber hub approaches $\bar{\Omega}^*_\theta$ for large $n$:
\begin{equation}\label{eq:12}
\Omega^*_{\text{II}} (n) = \bar{\Omega}^*_\theta, \quad \text{for}~n \to \infty. 
\end{equation}
There is no known analytical solution for $\bar{\Omega}^*_\theta$, and only numerical solutions are available, such as the envelope curve developed by Padday and Pitt \cite{padday1973stability}, the empirical models developed by Sadullah et al. \cite{sadullah2024predicting} and Daniel and Koh \cite{daniel2023droplet}. 

In the current work, we use the software Surface Evolver (SE, version 2.70) \cite{brakke1992surface} to simulate the profiles of droplets of increasing volume to determine $\bar{\Omega}^*_\theta$. 
The liquid phase with a specified volume is initialized as a cuboid-shaped slab hanging beneath a flat solid surface of a prescribed contact angle. 
SE then evolves the liquid cuboid towards its equilibrium shape by minimizing the total free energy (see Section~4 in ESI).
We carry out a series of simulations by increasing the volume of the droplet from a small value with an increment of \SI{1}{\micro\liter} until reaching a critical volume, at which a larger droplet detaches from the flat surface.

Figure~\ref{fig:5}(A) and (B) compare the simulated and actual profiles of the maximum \SI{0.01}{M} and \SI{0.001}{M} SDS droplets suspended beneath a flat surface.
The simulated profiles by SE closely resemble the image of droplets from experiments.
The maximum droplet volumes from simulations and experiments show good agreement.
As additional verifications, the model by Padday and Pitt \cite{padday1973stability} gives \SI{120}{\micro\liter} (calculated by $\bar{\Omega}^*_{\theta = 6^\circ} = 18.6\lambda^3$, see ESI Section 5) and \SI{96}{\micro\liter} ($\bar{\Omega}^*_{\theta = 48^\circ} =10.1\lambda^3$) for \SI{0.01}{M} and \SI{0.001}{M} SDS, respectively. 
The model by Daniel and Koh \cite{daniel2023droplet} gives \SI{118}{\micro\liter} (calculated by Eq. 4 in \cite{daniel2023droplet}) and \SI{102}{\micro\liter}, respectively. 
Table~\ref{tab.1} summarizes the cross validation of the maximum volume of SDS droplets from the experiments, SE-based simulations, and the results by Padday and Pitt \cite{padday1973stability}, and Daniel and Koh \cite{daniel2023droplet}.

\begin{table}[!h]
\centering
\caption{Cross validation of simulations and experiments for the maximum SDS droplets suspended beneath flat surfaces. The uncertainty (\SI{1}{\micro\liter}) for the Surface Evolver simulations is determined by the incremental volume used in the simulations.}
\label{tab.1}
\begin{tabular}{@{}lcccc@{}}
\toprule
\multirow{2}{*}{SDS concentration [M]} & \multicolumn{4}{c}{Maximum volume: $\Omega$ [\SI{}{\micro\liter}]}  \\ \cmidrule(l){2-5} 
                                   & Experiment& Simulation& Padday and Pitt~\cite{padday1973stability}& Daniel and Koh~\cite{daniel2023droplet}\\ \midrule
\multicolumn{1}{c}{0.01}           & $105\pm7$                & $110 \pm 1$               & 120                    & 118\\
\multicolumn{1}{c}{0.001}          & $93\pm4$                 & $92 \pm 1$               & 96                    & 102\\ \bottomrule
\end{tabular}
\end{table}

\subsection*{Validation and Discussion}

\begin{figure}[!h]
    \centering
    \includegraphics[width=1\linewidth]{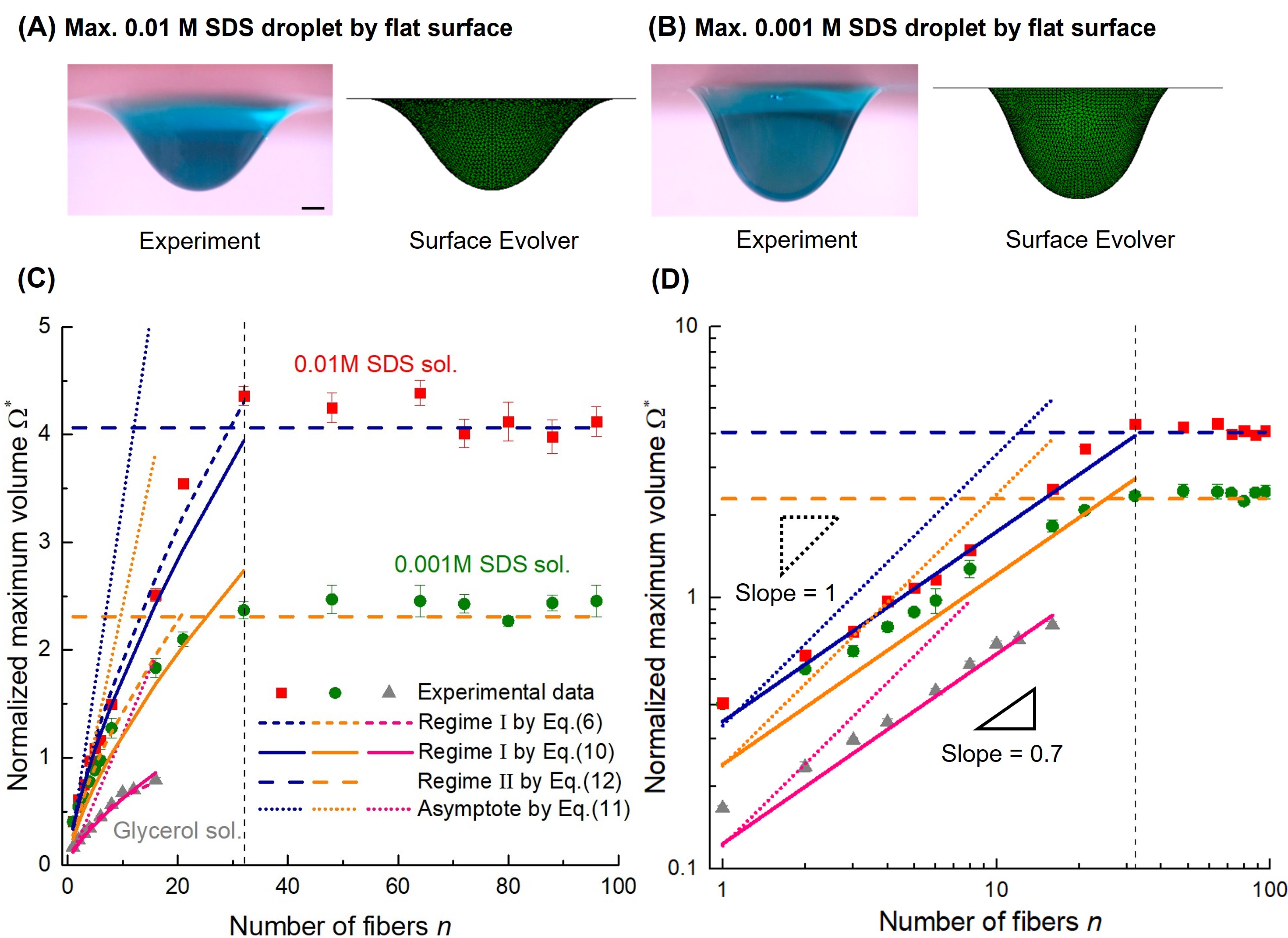}
    \caption{The experimental and simulated profiles of the maximum (A) 0.01~M and (B) 0.001~M SDS droplets hanging beneath flat surfaces (scale bar indicates \SI{1}{mm}). (C) Validation of the analytical model \eqref{eq:6}, the semi-empirical model \eqref{eq:10}, and the asymptote \eqref{eq:11} for regime I, and the model \eqref{eq:12} for regime II against experimental data (the vertical dashed line represents the critical fiber count $n^*= 32$, as also shown in (D)). (D) On a logarithmic scale of (C).} 
    \label{fig:5}
\end{figure}

In this section, we validates the models developed in the previous sections for regimes I and II  against the experimental data (see Figure~\ref{fig:5}(C)).
In regime I, the analytical model (\eqref{eq:6}, short dashed lines) and the semi-empirical model (\eqref{eq:10}, solid lines) predict the maximum droplet volume for different $n$, respectively, for the three types of solutions. 
Both models exhibit good agreement with the experimental data, with a mean relative error of 6.6\% and 10.8\% for the analytical and semi-empirical models, respectively.
In regime II, the model for the maximum volume (\eqref{eq:12}, dashed lines),  adopts the values from the SE-based simulations for $\bar{\Omega}_{\theta}^*$ (see Table~\ref{tab.1}). 
The model provides reasonable prediction of the maximum volume for regime II with a mean relative error of 4\%.

Figure~\ref{fig:5}(D) presents Figure~\ref{fig:5}(C) in logarithmic scale, highlighting the scaling of $\Omega^*$ with respect to $n$. 
In regime~I, the asymptote \eqref{eq:11} predicts the maximum volume by multiplying $\Omega^*_{\text{I}}(n=1)$ for a single fiber by $n$, yielding $\Omega^*  \sim n^1$, resulting in straight dotted lines with a slope of unity.
While the maximum volume from experiments, exhibiting strong linearity in logarithmic scale, follows straight solid lines with a slope of 0.7, which rationalizes the prediction by the semi-empirical model \eqref{eq:10}, i.e., $\Omega^*\sim n^{1}\tan\beta=n^{0.7}$ for $n \le 3n^*$.
The deviation of the experimental maximum volume ($\Omega^* \sim n^{1}\tan\beta$) from the asymptote ($\Omega^* \sim n^{1}$) arises from the reduced off-axis angle ($\tan\beta\sim n^{-0.3}$), leading to a lower growth rate of maximum volume as fiber count increases, see Figure~\ref{fig:2}(A\&D).
In regime~II, \eqref{eq:12} agrees with the experimental data and captures the scaling behavior of $\Omega^*  \sim n^0$ for $n>n^*$.
\eqref{eq:11} and \eqref{eq:12} can be viewed as two asymptotes for $\Omega^*(n)$, in the limits of $n\to1$ (dotted lines) and $n\to\infty$ (dashed lines in Figure \ref{fig:5}(C\&D)), respectively.

From Figure~{\ref{fig:5}}(C\&D), the influence of contact angle~($\theta$) on the normalized maximum volume~($\Omega^*_{\text{I}}$) in regime I is not straightforward. 
In general, increasing $\theta$ from $6^{\circ}$ ({\SI{0.01}{M}} SDS solution) to $48^{\circ}$ (0.001 M SDS solution) results in a decrease in $\Omega^*_{\text{I}}$. 
This trend is consistent with \eqref{eq:10}, where the term $(1-\theta/\pi)\cos\theta$ decreases as $\theta$ increases, thus reducing $\Omega^*_{\text{I}}$. 
However, the relationship is complicated by the fact that a larger $\theta$ can also increase the off-axis angle ($\beta_{n=1}$), see ESI Section~6.
This leads to in a greater $\tan\beta_{n=1}$, i.e., the value of $C$ in \eqref{eq:10}, potentially offsetting the decrease from the $(1-\theta/\pi)\cos\theta$ term. 
Since $\beta$ is typically measured by experiment \cite{lorenceau2004capturing, pan2018drop}, directly predicting the exact influence of $\theta$ on $\Omega^*_{\text{I}}$ without measuring $\beta$ remains challenging. 
Notably, a moderate $\theta$ of $30^{\circ}$ (glycerol solution) yields an even lower $\Omega^*_{\text{I}}$ than both SDS solutions in our experiments. 
This is attributed to the much smaller normalized fiber radius ($r^*= 0.02$) in the glycerol case, compared to $>0.12$ for the SDS droplets on resin fibers.
In regime II, increasing $\theta$ consistently decreases the normalized maximum volume ($\Omega^*_{\text{II}}$), a trend supported by existing models \cite{padday1973stability, daniel2023droplet} and confirmed by our work.

For the critical fiber number ($n^*$), the influence of $\theta$ on $n^*$ can be analyzed based on the models for regimes I and II. 
$\Omega^*_{\text{I}}=\Omega^*_{\text{II}}$ holds at $n=n^*$, so \eqref{eq:12} can be substituted into \eqref{eq:10} to achieve
\begin{equation}\label{eq:13}
\bar{\Omega}^*_\theta \approx 3C (n^*)^{0.7}r^*\left(1-\frac{\theta}{\pi}\right)\cos\theta.
\end{equation}
Isolating $n^*$ from \eqref{eq:13} provides a prediction of the critical fiber count.

As discussed above, increasing $\theta$ reduces both $\bar{\Omega}^*_\theta$ and the term $(1-\theta/\pi)\cos\theta$, but increases the coefficient $C$. 
By analyzing the variation of $\bar{\Omega}^*_\theta$ with respect to $\theta$ using existing models~\cite{padday1973stability}, we find that the ratio of $\bar{\Omega}^*_\theta$ to $(1-\theta/\pi)\cos\theta$ remains roughly constant at around 4.5 for $\theta<50^{\circ}$ (see ESI Section 7 for details). 
Consequently, the increase in $C$ due to the larger $\beta_{n=1}$ leads to a smaller $n^*$.
In addition, the ratio 
\begin{equation}
\label{eq:14}
    \frac{\bar{\Omega}^*_\theta}{(1-\theta/\pi)\cos\theta} \approx 4.5
\end{equation}
allows a simplified formula to predict $n^*$. Combining \eqref{eq:14} and \eqref{eq:13} leads to 
\begin{equation}
\label{eq:15}
    n^* \approx \left(\frac{1.5}{Cr^*}\right)^{1.4},
\end{equation}
where $1.4$ is derived from $1/0.7$.
{\eqref{eq:15}} is practically useful as, for small $\theta$, $n^*$ can be roughly predicted by knowing $r^*$ and a single experiment measuring $C=\tan(\beta_{n=1})$ without measuring contact angle.

Based on \eqref{eq:13}, the theoretical $n^*$ for the \SI{0.01}{M} and \SI{0.001}{M} SDS solutions are calculated to be 35 and 27, respectively, matching well with the experimental results (similar conclusion can be drawn using \eqref{eq:15}). 
Although $\theta$ increases from $6^{\circ}$ to $48^{\circ}$, the corresponding variation of $n^*$ is relatively modest. 
The rational of this observation lies in that the increase in $\tan\beta_{n=1}$ is partially offset by the decreased $r^*$ since there is an increase in capillary length ($\lambda$) when increasing $\theta$.
Assuming that the changes of $\theta$ ($5^{\circ}-50^{\circ}$) and $\lambda$ ($1.7-2.7$~mm) are independent to each other, $n^*$ should range from $\sim25$ to $\sim53$ based on \eqref{eq:15}. 
For the glycerol solution on 0.1~mm nylon fibers, its $n^*$, although not directly measured by experiments, is expected to be much larger. 
A theoretical $n^*$ of 108 is computed by \eqref{eq:15}, which appears reasonable by extrapolating the data points for glycerol solution in Figure~\ref{fig:5}(D) to a plateau $\Omega^*_{\text{II}}$ of $\sim3.3$ (obtained from Figure~S4). In this case, the effect of $r^*$ becomes the dominant factor.
In summary, the specific critical value $n^*=32$ is not universal.
However, we believe that the existence of a critical fiber count as a result of the competition between the mechanisms governing the regimes I and II holds as a more universal argument.

\section*{Conclusion}
\label{sec: Conclusion}
The current work studied the capability of fiber hubs, consisting of $n= 1$ to 96 fibers, to retain liquid droplets under them.
Analytical and semi-empirical models are developed and validated against experimental data. 
The stability of a droplet and its maximum volume suspended under a fiber hub vary across two distinct regimes, separated by a critical fiber count ($n^*$).
In regime I ($n\le n^*$), the stability of droplets is determined by the pinning of three-phase contact lines, and the maximum volume increases with fiber numbers. 
However, the growth rate is discounted by the reducing off-axis angle, due to the enhanced spreading of droplets along the fiber. 
In regime II ($n>n^*$), the maximum volume plateaus with respect to the number of fibers, and the instability is mainly driven by the necking of the droplet.
Before detaching, the shape of a droplet suspended beneath a fiber hub closely resembles that of a droplet under a flat surface, with its stability governed by Rayleigh-Taylor instability.

\section*{Support Information}
\begin{itemize}[itemsep=0pt]
  \item Measurement of the maximum volume of the droplet
  \item Identification of the center of mass of a droplet
  \item Approximation of Eq~(9) in the Main Text
  \item Expression for total free energy
  \item Extracting data from Padday and Pitt (1973)
  \item Influence of contact angle on the off-axis angle
  \item Variation of the normalized maximum volume of regime II
\end{itemize}

\section*{Acknowledgement}
The authors express their sincere gratitude for the financial support provided by the Artificial Intelligence for Design Challenge (AI4D) program, funded by the National Research Council Canada. 

\bibliographystyle{achemso}

\bibliography{references1}

%

\end{document}